\documentclass[reviewcopy]{elsart}

\usepackage{adnd}
\usepackage{longtable}

\usepackage{mathptmx}

\usepackage{amsmath}

\usepackage{amssymb}

\usepackage{graphicx}

\begin{document}

\begin{frontmatter}

\journal{Atomic Data and Nuclear Data Tables}

\copyrightholder{Elsevier Science}

\runtitle{Optimized effective potential atomic wave functions}
\runauthor{Sarsa, Buend\'{\i}a and G\'alvez}


\title{
Parameterized optimized effective potential for the ground state
of the atoms He through Xe
}

\author{
A. Sarsa, F. J. G\'alvez}
\author{
\MakeLowercase{and}
E. Buend\'{\i}a}
\address{
Departamento de F\'{\i}sica Moderna, Facultad de Ciencias,\\
Universidad de Granada,  E-18071 Granada, Spain}

\date{03.12.2003} 

\begin{abstract}  
Parameterized orbitals expressed in Slater-type basis obtained within
the optimized effective potential framework as well as the parameterization
of the potential are reported for the ground state of the atoms He through Xe.
The total, kinetic, exchange and single particle energies are given for 
each atom.
\end{abstract}

\end{frontmatter}





\newpage
\tableofcontents
\listofDtables

\vskip5pc


\section{INTRODUCTION}

In this work we report parameterized atomic wave functions obtained within 
the optimized effective potential (OEP) framework for the $2 \leq Z \leq 54$
atoms.
The optimized central effective potential is also given in a parameterized 
form.
The radial part of the atomic orbitals is expanded in a basis set of
Slater-type functions as in the Roothaan-Hartree-Fock method
\cite{clro74,bbb93}. 
The effective potential is parameterized in terms of yukawian functions
times powers of $r$ in such a way that the
asymptotic exact behavior is fulfilled and all the integrals involved 
can be calculated analytically.

The optimized effective potential method has its origin on the
approximate X-$\alpha$ Slater's simplified treatment of the exchange term
in the Hartree-Fock equations based on
the averaging in the occupied orbitals \cite{slater51}. 
The method was formulated 
by Sharp and Horton \cite{shho53} in the independent particle model
as a variational alternative to the previous scheme.
As in the Hartree-Fock approach, the starting point is the
expectation value of the electrostatic $N$-electron
Hamiltonian in a single Slater determinant or
in a single configuration in the $L$, $S$ coupling.
In the OEP method an additional constraint is imposed into the variational
problem with respect to the Hartree-Fock one: the orbitals must satisfy a 
single-particle Schr\"odinger
equation with a certain local potential, the same for all the
electrons (with the same spin). Then the expectation value of the
Hamiltonian of the $N$-electron system becomes a functional of such a
local potential. The effective potential is then varied to minimize
the total energy. This gives rise to a linear
integral equation in the effective potential
\cite{shho53,tash76} whose solution provides the optimized effective
potential.  
This equation was first solved by Talman and coworkers by using a
numerical scheme \cite{tash76} and a complete tabulation for the
ground state of the atoms Li through Rd was reported \cite{alt78}.

However the use and application of the OEP method was hindered by
the computational difficulties posed by the integral equations.
An alternative methodology that simplifies greatly these problems
giving rise to a better accuracy of the solution is based in a
finite basis set expansion and/or analytical forms for the effective
potential \cite{dabe81,lasset85,fryu98,cone01,yawu02,sgb03}.
In this work we
apply systematically a recently proposed scheme \cite{sgb03} within
the parameterized methodology for the ground state configuration of the
He through Xe atoms.  
For each atom the total, kinetic, exchange and single particle energies 
are given.
Some analytical constraints, that must be fulfilled by the exact solution,
are imposed in the variational search.

\subsection{The Effective potential method}

The orbitals and effective potentials tabulated in this work have been
obtained by using the method of Ref. \cite{sgb03}. Below we give a brief
overview of the effective potential method and the parameterization 
scheme followed.

The OEP method is a variational approximation to the many electron problem
which uses single-particle wave functions as variational ansatz as
in the Hartree-Fock method. 
The single particle orbitals satisfy a single
particle Schr\"odinger equation with a certain local potential that
is taken to be central
\begin{equation}
\label{schro}
\left(-\frac{1}{2}\vec{\nabla}^2 + V_{e}(r)\right) \, 
   \phi_{\lambda,\sigma}(\vec{r}) = 
\epsilon_{\lambda} \phi_{\lambda,\sigma}(\vec{r})
\end{equation}
where $\lambda=(n, l, m_l)$  stands for the spatial quantum numbers
of the atomic orbital and
$\sigma=\pm 1/2$ is the third component of the electron spin.

The expectation value of the atomic Hamiltonian is then a functional
of this effective potential.
The minimum condition on the energy gives rise to a linear
integral equation on the effective potential
\cite{shho53,tash76} whose solution provides the optimized effective
potential.  
The energy so obtained is an upper bound to the exact one and is
above the Hartree-Fock value. 
It is also worth pointing out 
that the role of the effective potential here is just an auxiliary function
used to calculate the orbitals in the wave function.

Several analytical conditions are known for the effective potential.
First its short- and long- range asymptotic behavior
\begin{equation}
\label{asymp}
V_e(r) \longrightarrow -\frac{Z}{r} ~~~  
\mbox{as } r \rightarrow 0 ; ~~~~~~
V_e(r) \longrightarrow -\frac{Z-(N-1)}{r} ~~~  
\mbox{as } r \rightarrow \infty 
\end{equation}

Second the exact solution of the OEP must
also satisfy the virial relation and  the exchange--only
virial relation  \cite{ghpa85,lepe85}
\begin{equation}
E_x=E_x^{vr}
\end{equation}
where  $E_x^{vr}$ is given by 
\begin{equation}
E_x^{vr} = 
- \int {\rm d} \vec{r} \rho(\vec{r}) \vec{r} \cdot \vec{\nabla} 
    V_x(\vec{r}) 
~~; ~~~~~~~~
V_x(\vec{r})=V_e(r)+\frac{Z}{r}-
\int {\rm d} \vec{r}^{\, \prime} \, 
\frac{\rho(\vec{r}^{\, \prime})}
     {\mid \vec{r} -\vec{r}^{\, \prime} \mid}
\end{equation}
and $V_x$ is the exchange potential and $E_x$ is the usual exchange
energy
\begin{equation}
E_{x} = -\frac{1}{2} \sum_{\sigma} \sum_{\lambda, \mu =1}^{N_{\sigma}}
\int {\rm d} \vec{r} \int {\rm d} \vec{r}^{\,\prime} \frac{\phi_{\lambda
\sigma}^\ast(\vec{r})\phi_{\mu \sigma}^\ast({\vec{r}^{\,\prime}}) \phi_{\mu
\sigma}(\vec{r})\phi_{\lambda \sigma}({\vec{r}}^{\,\prime})}{\mid\vec{r} -
\vec{r}^{\,\prime}\mid}
\end{equation}

The third condition \cite{kli90} states that for the exact solution the 
eigenvalue of the highest occupied orbital in equation \eqnref{schro}
must coincide with the corresponding Hartree-Fock single particle
energy,
$\epsilon_{\lambda_h} = \epsilon^{\rm HF}_{\lambda_h}$,
where $\lambda_h$ stands for the spatial quantum numbers of the highest
occupied orbital and 
the Hartree-Fock single particle energies are computed as
\begin{equation}
\label{singleener}
\epsilon^{\rm HF}_{\lambda}
 = I_{\lambda} + 
\sum_{ \mu}
(J_{\lambda \mu} - K_{\lambda \mu})
\end{equation}

In this equation $I$, $J$ and $K$ are the usual
single particle, direct and exchange terms calculated starting from
the eigenfunctions of the effective potential.  

The last two conditions are in some sense complementary. 
The main contributions to the 
quantities involved in the exchange only virial condition arise 
from the internal region of the atoms whereas the highest energy 
eigenvalue is governed by the outermost region. 

\subsection{The Parameterized solution}

To minimize the expectation value of the atomic Hamiltonian with respect
to the effective potential both the effective potential and the
atomic orbitals have been expanded in an auxiliary basis set. Then
the total energy becomes a function of several variational parameters which
have been optimized by using the SIMPLEX algorithm \cite{numrcip91}.

The parameterization taken for the  effective potential is 
\begin{equation}
V_e(r)=-\frac{1}{r} \left( Z-N+1+(N-1) \sum_{p=0}^S
\sum_{k=1}^{n_p} c_{k,p} r^p e^{-\beta_{k,p} r} \right)
\label{paref}
\end{equation}
with the condition
\begin{equation}
\sum_{k=1}^{n_0} c_{k,0}=1
\label{eq.constraint}
\end{equation}
imposed in order to match the correct short-range behavior of the
potential. This constraint makes that we must deal with, at least, two
terms with $p=0$. 
With this analytical dependence the correct long range behavior is 
also fulfilled.

The spatial part of the atomic orbitals is written in terms of the
spherical harmonics 
\begin{equation}
\phi_\lambda(\vec{r})=R_{n\,l}(r) Y_{l\, m_l}(\Omega)
\end{equation}
and the radial atomic orbitals $R_{nl}$ have been expanded in terms of 
Slater-type basis functions
\begin{equation}
R_{n\,l}(r)=\sum_{j=1}^{M_l} C_{j\,l\,n} \, S_{j\,l}(r) 
\label{radial}
\end{equation}
where $S_{j\,l}(r)$ is given by
\begin{equation}
S_{j\,l}(r)=N_{j\,l} ~ r^{n_{j\,l}-1} \, e^{\alpha_{j\,l} r}
\label{sto}
\end{equation}
and the normalization constant $N_{j\,l}$ is
\begin{equation}
N_{j\,l}=
\frac{(2 \, \alpha_{j\,l})^{n_{j\,l}+1/2}}
{\sqrt{(2 \,n_{j\,l})!}}
\end{equation}

\subsubsection{Algorithm}

A set of basis functions must be fixed for both the  radial part of 
the atomic orbitals and the  effective potential.
For the Slater-type orbitals we have used the same values of
$M_l$ and $\{n_{j\,l}\}$ in equations \eqnref{radial} and \eqnref{sto}
respectively as in the Roothaan-Hartree-Fock solution of
Ref. \cite{bbb93}.
With respect to the parameterization of the effective potential the
number of basis functions used to expand the effective potential is
incremented systematically until convergence is reached.

Once the expansion for the effective potential and atomic orbitals
is fixed, the both sets of free parameters
$\{ c_{k \, p}, \beta_{k \, p} \}$ for
the potential, and $\{ n_{j \, l}, \alpha_{j \, l} \}$ for the orbitals
are initialized to some given value. 
Then the single particle Schr\"odinger equation \eqnref{schro} is solved 
in the basis given by the Slater type orbitals selected. This is the usual
generalized eigenvalue problem
\begin{equation}
\label{generalized}
{\bf H}(l) \cdot \vec{C}_{l \,n}= 
\epsilon_{l \, n} ~~ {\bf N}(l) \cdot \vec{C}_{l \,n}
\end{equation}
where
\begin{equation}
\vec{C}_{l \,n}= \{C_{j\, l \,n} \} ~~~~
\mbox{with} ~~~~ j=1,\ldots, M_l
\end{equation}
and the matrices ${\bf H}(l)$ and ${\bf N}(l)$ are given by
\begin{equation}
({\bf H}(l))_{j \, k} =
\langle S_{j\, l} \mid 
\left(-\frac{1}{2}\vec{\nabla}^2 + V_{e}(r)\right) 
\mid S_{k\, l} \rangle ,  ~~~
({\bf N}(l))_{j \, k} =
\langle S_{j\, l} \mid S_{k\, l} \rangle 
\end{equation}

The solution of equation \eqnref{generalized} is obtained by using the usual 
linear algebra algorithms and  provides the eigenvalues and the linear 
coefficients $\{C_{j\, l \,n} \}$ $j=1,\ldots, M_l$. 
Then the atomic orbitals are fully determined.

The $N$-electron wave function, $\Phi$, is a linear combination of Slater
determinants with coefficients fixed for each $L$, $S$ symmetry. The orbitals
have been determined in the previous step and therefore
the expectation value of the non relativistic atomic Hamiltonian can be 
computed readily
\begin{equation}
\langle H \rangle= 
\langle \Phi \mid 
-\frac{1}{2}\sum_{i=1}^N 
\left(\vec{\nabla}_i^2 +\frac{Z}{r_i} \right)
\mid \Phi \rangle 
+\langle \Phi \mid 
\sum_{i<j}^N \frac{1}{r_{i\,j}} 
\mid \Phi \rangle 
\end{equation}

This can be reduced to one and two body integrals that 
can be computed analytically because we are using Slater-type orbitals.

As a result,
the expectation value of the Hamiltonian becomes a function of the non linear
parameters $\{ c_{k \, p}, \beta_{k \, p} \}$ of the effective potential, and
$\{\alpha_{j \, l} \}$ of the radial orbitals. 
The following functional 
\begin{equation}
\langle H \rangle + \mid 
                    \epsilon_{\lambda_h} - \epsilon^{\rm HF}_{\lambda_h}
		    \mid
+ \mid E_x-E_x^{vr}\mid
\end{equation}
has been minimized with respect to these parameters in order to fulfill
the analytic conditions of the exact effective potential.
As a result both of them hold within $10^{-6}$ hartree 
and the minimum energy is not substantially different to that obtained 
in an unconstrained search. 
The tail of the effective potential is expected to 
be better reproduced by imposing the homo-condition.

The quality of the results here reported can be visualized in 
Figure A where we plot the relative error
(in $\%$) of our ground state energy  with respect to the HF one 
\cite{bbb93} as
compared to the relative error of the numerical solution \cite{alt78}.
In the present work the relative error is
nearly constant for all the atoms considered.
It is also noticeable the great improvement of the parameterized solution with
respect to previous solutions especially for the lighter atoms.

\ack
This work has been partially supported by the Ministerio de Ciencia y
Tecnolog\'{\i}a and FEDER under contract  BFM2002-00200,  
and by the Junta de Andaluc\'{\i}a.

\section*{FIGURES}

\begin{figure}[ht!]
\label{figure1}
\caption{
Relative difference (in percent) between the OEP ground state energy
and the HF one for the atoms He through Xe. 
The parameterized results of this work are compared with the
numerical ones of \cite{alt78}.
}
\end{figure}

\newpage

\setcounter{figure}{0}

\begin{figure}[ht!]
\begin{center}
\includegraphics{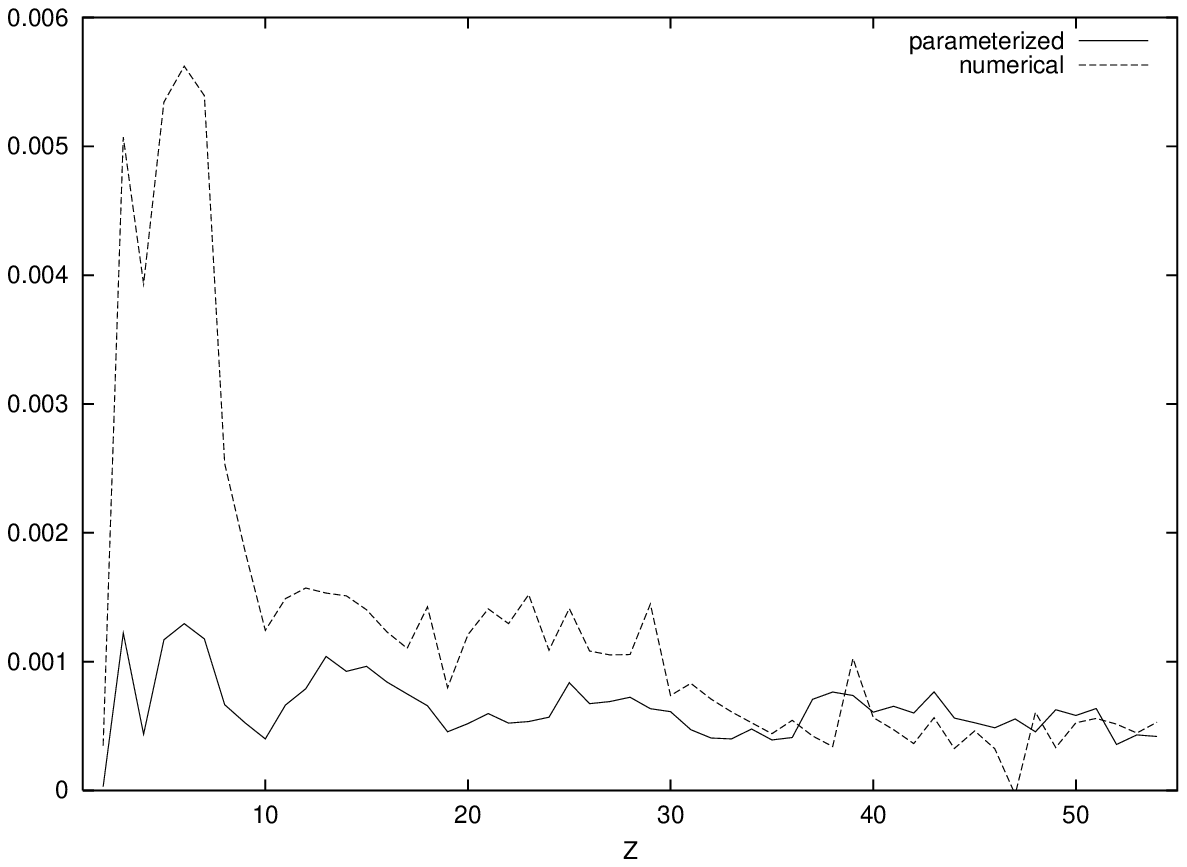}
\caption{
}
\end{center}
\end{figure}

\newpage

\section*{EXPLANATION OF TABLE}\label{sec.eot}
\addcontentsline{toc}{section}{EXPLANATION OF TABLE}

The first row of each data block gives the name of the element,
the atomic number, the single particle configuration and the
term of the ground state wave function. 
The second row gives the total, kinetic and exchange energy.
Then the parameterization of both the effective potential and the
atomic orbitals is given.
Atomic units are used.

\renewcommand{\arraystretch}{1.0}

\begin{tabular}{@{}p{1in}p{6in}@{}}
\textbf{TABLE.} & \textbf{Parameterized atomic wave function and
                          effective potential 
		from \textit{Z}=2 to \textit{Z}=54.}\\[8pt]
$Z$		& Atomic number \\
$1s^2$, etc.	& Atomic shell and electron occupation number \\
                & Parameterization of the effective potential\\
Column 1        & The powers in the parameterization of
                  the potential in equation \eqnref{paref} \\
Column 2        & The exponents in the parameterization of 
                  the potential in equation \eqnref{paref} \\
Column 3        & The linear coefficients in the parameterization of 
                  the potential in equation \eqnref{paref} \\
                & Parameterization of the orbitals \\
$1S, 2S$, etc 	& The values $1$, $2$ gives $n_{j\,l}$ 
                   in equation \eqnref{sto} \\
Column 1        & The values of the exponent  $\alpha_{j\,l}$
		  in equation \eqnref{sto} \\
Other columns   & The first entry is the Hartree-Fock orbital energy
                  $\epsilon^{\rm HF}_{\lambda}$
		   in equation \eqnref{singleener}\\
                & The second entry is the single particle energy
                  $\epsilon_{\lambda}$
		  in equation \eqnref{schro} \\
                & The other entries are the coefficients $C_{j\,n\,l}$
		  in equation \eqnref{radial} \\
\end{tabular}
\label{table}

\newpage
\datatables

\setlength{\LTleft}{0pt}
\setlength{\LTright}{0pt} 

\renewcommand{\arraystretch}{1.0}

\footnotesize
\begin{longtable}[c]{%
@{}@{\extracolsep{\fill}}r@{\hspace{5\tabcolsep}}rrrrrrr@{}}
\caption[Atomic wave functions and effective potential]%
{Parameterized atomic wave function and effective potential
	from $Z=2$ to $Z=54$}\\[0pt]
\caption*{\small{See page \pageref{table} for Explanation of Tables}}\\
\hline
\\[100pt]
\multicolumn{7}{c}{\textit{This space intentionally left blank}}\\
\endfirsthead
\\ 
\multicolumn{5}{l}{
HELIUM     \quad $Z=$ 2 \quad 1s$^{2}$                       \quad $^1$S}\\[2pt]
\multicolumn{2}{c}{TOTAL    ENERGY} &
\multicolumn{3}{c}{KINETIC  ENERGY} &
\multicolumn{2}{c}{EXCHANGE ENERGY} \\
\multicolumn{2}{c}{$     -2.86167914$}&
\multicolumn{3}{c}{$      2.86167914$}&
\multicolumn{2}{c}{$     -1.02580341$}\\[2pt]
\multicolumn{2}{l}{PARAMETERS VEFF}\\
\multicolumn{1}{c}{$p$} & \multicolumn{1}{c}{$\beta_{k\, p}$} &
\multicolumn{1}{c}{$c_{k\, p}$}\\
\multicolumn{1}{c}{2}  & $      0.81080622$ & $     -0.33200330$\\
\multicolumn{1}{c}{2}  & $      3.64179068$ & $     -1.10884184$\\
\end{longtable}

\end{document}